\documentclass[pre,aps,reprint,superscriptaddress]{revtex4-1}
\usepackage{graphics}
\usepackage{graphicx}
\usepackage{amsmath}
\usepackage{float}
\usepackage{xcolor}

\begin{document}

\title{Shear Is Not Always Simple: Rate-Dependent Effects of Flow Type on Granular Rheology} 

\author{Joel T. Clemmer}
\affiliation{Sandia National Laboratories, Albuquerque, New Mexico 87185, USA}
\author{Ishan Srivastava}
\affiliation{Center for Computational Sciences and Engineering, Lawrence Berkeley National Laboratory, Berkeley, California 94720, USA}
\author{Gary S. Grest}
\affiliation{Sandia National Laboratories, Albuquerque, New Mexico 87185, USA}
\author{Jeremy B. Lechman}
\affiliation{Sandia National Laboratories, Albuquerque, New Mexico 87185, USA}
\date{\today}

\begin{abstract}

Despite there being an infinite variety of types of flow, most rheological studies focus on a single type such as simple shear.
Using discrete element simulations, we explore bulk granular systems in a wide range of flow types at large strains and characterize invariants of the stress tensor for different inertial numbers and interparticle friction coefficients.
We identify a strong dependence on the type of flow which grows with increasing inertial number or friction.
Standard models of yielding, repurposed to describe the dependence of the stress on flow type in steady-state flow and at finite rates, are compared with data.

\end{abstract}

\maketitle

%%%%%%%%%%%%%%%%%%%%%%%%%%%%%%%%%%%%%%%%%%%%%%%%%%%%%%%%%%%%
%                      INTRODUCTION
%%%%%%%%%%%%%%%%%%%%%%%%%%%%%%%%%%%%%%%%%%%%%%%%%%%%%%%%%%%%

The rheology of many materials is often characterized by relating the shear stress to the strain rate through a scalar viscosity. 
However, the strain-rate tensor $\dot{E}_{ij}$ and the stress tensor $T_{ij}$ contain more information on the flow of the system and their connection cannot necessarily be reduced to such a scalar relation as it may be more complex \cite{Schunk1990,Giusteri2018}.
For instance, their eigenvalues may not be related by a single proportionality constant.
To fully characterize the tensorial relationship between $\dot{E}_{ij}$ and $T_{ij}$, one must also explore more than one type of flow.
In dense granular gases, significantly different behavior is observed in laminar versus shear flows \cite{Reyes2011, Wu2015}.

While studies often focus on planar shear flow, there exists a continuous spectrum of types of flow.
The type of flow has been found to have a significant effect on yield strength \cite{Thornton2010}, dilatancy \cite{Goddard1998}, and fluctuations \cite{Didwania2001} in granular materials and there is a need to develop and apply new methods to explore its effect on granular rheology.
Here we measure $T_{ij}$ in steady-state flow and focus on its alignment with $\dot{E}_{ij}$ and the dependence of a scalar shear stress [Eq. \eqref{eq:shear}] on the type of flow.
This dependence or stress envelope is a surface in the three-dimensional space created by the principal stresses.
It defines the set of possible principal stresses produced by a steady-state flow at a given inertial number, a dimensionless measure of strain rate defined in Eq. \eqref{eq:in}.

The dependence of shear stress on flow type at the onset of flow in granular materials is often described using simple models such as Mohr-Coulomb or Drucker-Prager \cite{Drucker1952}.
While these models traditionally characterize the yield stress in the small-strain limit \cite{Nedderman1992}, they can be extended to describe steady-state flows at large strains as in the original $\mu(I)$ model \cite{Jop2006}.
Despite frequent use, these simple models are inaccurate \cite{Thornton2010}, and advanced models are needed to capture the shape of stress envelopes \cite{Matsuoka1974, William1975, Lade1975}.

In this work, we use discrete element method (DEM) simulations to explore frictional granular rheology across different irrotational flow types, along with the well-studied simple shear flow.
Of these flows, only simple and pure shear are planar while the others are triaxial flows.
To reach large strains, we leverage generalized Kraynik-Reinelt boundary conditions \cite{Kraynik1992, Dobson2014, Hunt2016} which have been impactful in exploring the rheology of soft materials and complex fluids \cite{OConnor2018, OConnor2020}.
This work extends results from earlier studies on the dependence of granular rheology on flow type \cite{Thornton2010,Huang2014,Fleischmann2014,Cheal2018,Redaelli2019} to steady-state flows where we quantify the shape of the stress envelope for a wide range of inertial numbers and friction coefficients.

The relative importance of the flow type is measured in terms of a strength ratio $\Psi$, the ratio of the shear stress in triaxial extension (TXE) to compression (TXC).
Importantly, we find that $\Psi$, and therefore the shape of the envelope, heavily depends on the friction coefficient and inertial number.
As either of these parameters increase, $\Psi$ decreases as the type of flow has a greater impact on rheology.
Such characterizations are important for both fundamentally understanding the physics of flow and for developing tensorial formulations of granular rheology relating $\dot{E}_{ij}$ and $T_{ij}$ \cite{Weinhart2013,Giusteri2018,Srivastava2020b,Coquand2021} similar to the development of microstructure-aware constitutive models of suspensions \cite{Goddard2014}.
Our results are also used to assess the application of yield models to steady-state flow.

%%%%%%%%%%%%%%%%%%%%%%%%%%%%%%%%%%%%%%%%%%%%%%%%%%%%%%%%%%%%
%                      DEFINITIONS
%%%%%%%%%%%%%%%%%%%%%%%%%%%%%%%%%%%%%%%%%%%%%%%%%%%%%%%%%%%%

It is convenient to use the deviatoric stress and strain-rate tensors $\sigma_{ij} = T_{ij} + PI_{ij}$ and $\dot{\epsilon}_{ij} = \dot{E}_{ij} - 1/3 \dot{E}_V I_{ij}$ where $P$ is the pressure, $\dot{E}_V$ is the volumetric strain rate, and $I_{ij}$ is the identity tensor.
The eigenvalues of $\sigma_{ij}$ and $\dot{\epsilon}_{ij}$, or their principal components, are designated as $\sigma_1 \ge \sigma_2 \ge \sigma_3$ and $\dot{\epsilon}_1 \ge \dot{\epsilon}_2 \ge \dot{\epsilon}_3$.
In granular flows, $\sigma_{ij}$ and $\dot{\epsilon}_{ij}$ are approximately coaxial as they have equivalent time-averaged eigenvectors seen here and in Refs. \cite{Nedderman1992,Rycroft2009,Weinhart2013,Bhateja2020}.
However, their eigenvalues are not always simply proportional implying the tensors are not codirectional \cite{Silbert2001,Depken2007,Rycroft2009,Weinhart2013,Srivastava2020b} as further discussed below.

\begin{figure}
	\begin{center}
	\includegraphics[width=0.48\textwidth]{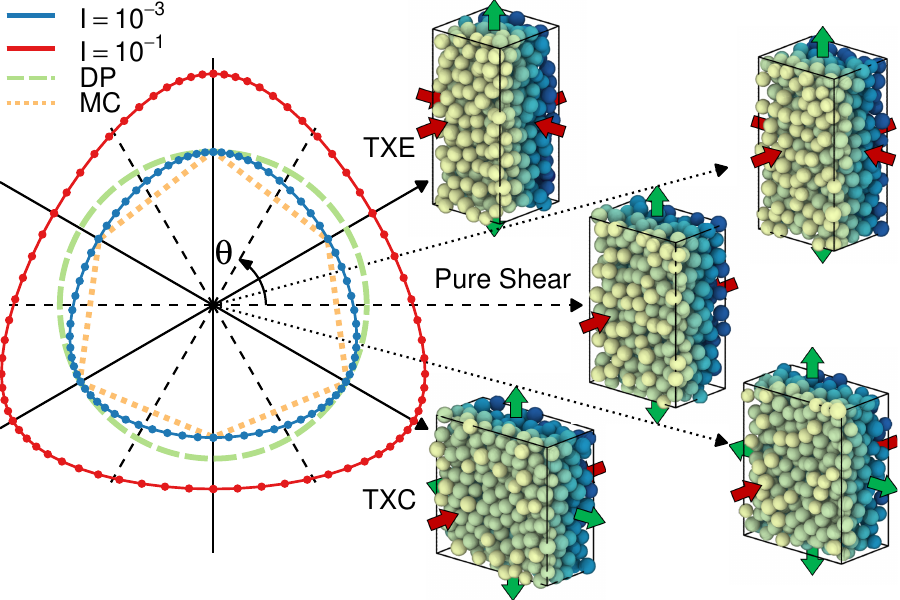}
	\caption{Constant-pressure profile of the average steady-state stress ratio $\mu$ (radial distance) as a function of Lode angle $\theta^\sigma$ (counterclockwise polar angle) for systems at inertial numbers $I$ of $10^{-3}$ (inner, blue) and $10^{-1}$ (outer, red) and interparticle friction $\mu_S = 0.1$. Predictions from Mohr-Coulomb (dotted, orange) and Drucker-Prager (dashed, green) models are included. Small, initially cubic systems are rendered after shearing for a strain $\epsilon_S = 0.5$ at $\theta^{\dot{\epsilon}} = -\pi/6$ (TXC), $-\pi/12$, $0$ (pure shear), $\pi/12$, and $\pi/6$ (TXE) to illustrate different types of flow using conventional boundary conditions. Arrows indicate contracted (red) and expanded (green) axes.} 
	\label{fig:example}
	\end{center}
\end{figure}

To quantify the magnitude of shear stress, we use the second invariant of $\sigma_{ij}$,
\begin{equation}
J^\sigma_2 = \frac{1}{2} \left( \sigma_1^2 + \sigma_2^2 + \sigma_3^2 \right) \ \ ,
\end{equation}
to define 
\begin{equation}
\sigma_S \equiv \sqrt{J^\sigma_2}\ \ .
\label{eq:shear}
\end{equation}  
The third invariant,
\begin{equation}
J^\sigma_3 = \frac{1}{3} \left( \sigma_1^3 + \sigma_2^3 + \sigma_3^3 \right) \ \ ,
\end{equation}
includes additional information on the direction of flow and is therefore used to define a Lode angle
\begin{equation}
\theta^\sigma = \frac{1}{3} \arcsin \left[ \frac{J^\sigma_3}{2} \left(\frac{3}{J^\sigma_2} \right)^{3/2}\right]
\end{equation}
commonly used to categorize types of flow \cite{Anand2020}.
In this definition, $\theta^\sigma$ varies between $-\pi/6$ and $\pi/6$ with the extreme cases corresponding to triaxial compression (TXC, $\sigma_2/\sigma_1 = 1$) and extension (TXE, $\sigma_2/\sigma_3 = 1$), respectively. 
Between these limits, a continuous spectrum of flows exists with a midpoint of $\theta^\sigma = 0$ corresponding to pure shear ($\sigma_2 = 0$).
Equivalent invariants are defined for the strain-rate tensor $\dot{\epsilon}_{ij}$: $J^{\dot{\epsilon}}_2$, $\dot{\epsilon}_S$, $J^{\dot{\epsilon}}_3$, and $\theta^{\dot{\epsilon}}$.
If $\theta^{\dot{\epsilon}} = \theta^{\sigma}$ then $\sigma_{ij}$ and $\dot{\epsilon}_{ij}$ are codirectional.
Systems sheared with different values of $\theta^{\dot{\epsilon}}$ are rendered in Fig. \ref{fig:example}.

%%%%%%%%%%%%%%%%%%%%%%%%%%%%%%%%%%%%%%%%%%%%%%%%%%%%%%%%%%%%
%                        METHODS
%%%%%%%%%%%%%%%%%%%%%%%%%%%%%%%%%%%%%%%%%%%%%%%%%%%%%%%%%%%%

Simulations of 40000 particles were run in LAMMPS \cite{Plimpton1995,Thompson2021} with diameters evenly distributed between $0.9$ and $1.1a$ and constant densities of $\rho = m/a^3$, where $a$ and $m$ are units of length and mass.
Similar to Refs. \cite{Srivastava2019,Srivastava2020a,Srivastava2020b}, interactions included Hookean normal forces with a stiffness $k$, equally stiff tangential frictional forces with a sliding friction coefficient $\mu_S$ \cite{Silbert2001}, and damping forces proportional to the difference in normal and tangential velocities with prefactors of 0.5 and 0.25, respectively, corresponding to a coefficient of restitution between $0.62$ and $0.75$, depending on particle radii \cite{Silbert2001}.
A velocity-Verlet integrator was used with a time step of $\Delta t = 0.02 \sqrt{m/k}$.
$T_{ij}$ included both kinetic and virial contributions \cite{Allen1989}.

Systems were initialized below jamming for all $\mu_S$ at a volume fraction of $\approx 0.52$ \cite{Silbert2010}.
The simulation cell was then deformed to maintain a fixed $\dot{\epsilon}_{ij}$ while affinely remapping particle positions. 
The principal components of $\dot{\epsilon}_{ij}$ are
\begin{align}
\begin{split}
\dot{\epsilon}_1 &= \dot{\epsilon}_S \left(\cos(\theta^{\dot{\epsilon}}) - \frac{1}{\sqrt{3}} \sin(\theta^{\dot{\epsilon}}) \right) \\
\dot{\epsilon}_2 &= \dot{\epsilon}_S \frac{2}{\sqrt{3}} \sin(\theta^{\dot{\epsilon}}) \\
\dot{\epsilon}_3 &= -\dot{\epsilon}_S \left(\cos(\theta^{\dot{\epsilon}}) + \frac{1}{\sqrt{3}} \sin(\theta^{\dot{\epsilon}}) \right)
\end{split}
\end{align}
using definitions of $\theta^{\dot{\epsilon}}$ and $\dot{\epsilon}_S$ and the fact that $\dot{\epsilon}_1 + \dot{\epsilon}_2 + \dot{\epsilon}_3 = 0$ since $\dot{\epsilon}_{ij}$ is trace free.
Generalized Kraynik-Reinelt periodic boundaries were used to reach large strains \cite{Dobson2014, Hunt2016, Nicholson2016}. 

To maintain a target pressure of $P_T = 10^{-5} k/a$, a Berendsen barostat was used to isotropically expand or contract the simulation cell \cite{Berendsen1984}.
The length $L$ of each side of the cell evolved according to 
\begin{equation}
L(t+\Delta t) = L(t) \left( 1 + \frac{P-P_T}{P_T}  \frac{\Delta t}{T_B} \right)^{1/3}
\label{eq:barostat}
\end{equation}
where $T_B$ is a damping time that controls how fast the barostat responds to deviations in pressure.
At small values of $T_B \lesssim T_c \equiv 0.1 \dot{\epsilon}_S^{-3/2} (k/m)^{1/4}$, $P$ is nearly constant while the volume $V$ fluctuates rapidly.
At $T_B \gtrsim T_c$, fluctuations in $P$ grow while $V$ stabilizes.
While fluctuations vary considerably with $T_B$, no significant effect was detected on the time-averaged stress tensor \footnote{A subset of data was collected for values of $T_B / T_c = 0.2$ and $10$ and no detectable effect was observed on any results in this Letter}.
In steady state, the time-averaged $V$ is constant and $\langle \dot{E}_V \rangle = 0$.

Initially using $T_B = 0.2 \sqrt{m/k}$ to accelerate compression, systems were sheared either to a strain of $\epsilon_S = $ 0.5 or for a duration of $10^5 \sqrt{m/k}$, whichever is longer, where $\epsilon_S = \dot{\epsilon}_S T$ and $T$ is the time sheared.
$T_B$ was then reduced to $0.1 T_c$, to minimize fluctuations in $P$ at all studied rates.
Above a strain of $1.0$, the system is in steady-state flow and average properties do not depend on strain \cite{Noll1962}. 
A stress ratio $\mu \equiv \sigma_S/P$ and $\theta^\sigma$ were calculated using a strain-averaged stress tensor for each combination of $\mu_S$, $\dot{\epsilon}_S$, and $\theta^{\dot{\epsilon}}$.

%%%%%%%%%%%%%%%%%%%%%%%%%%%%%%%%%%%%%%%%%%%%%%%%%%%%%%%%%%%%
%                        MU of I
%%%%%%%%%%%%%%%%%%%%%%%%%%%%%%%%%%%%%%%%%%%%%%%%%%%%%%%%%%%%

\begin{figure}
	\begin{center}
	\includegraphics[width=0.48\textwidth]{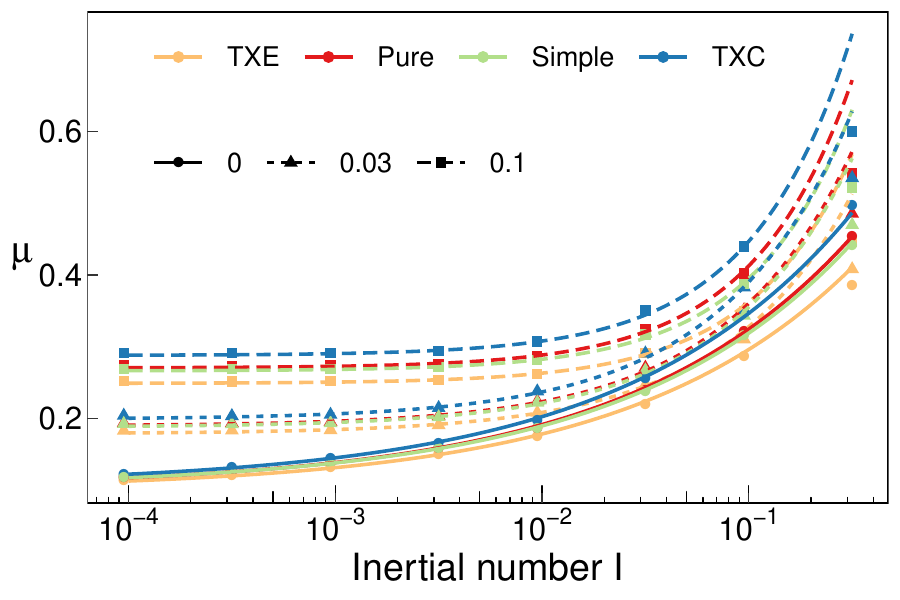}
	\caption{$\mu(I)$ curves for the indicated values of $\mu_S$ (line type and shape) in TXE (orange), pure shear (red), simple shear (green), and TXC (blue).
	Fits from Eq. \eqref{eq:HB} are overlaid with exponents of $0.4$, $0.7$, and $0.9$ for $\mu_S = 0.0$, $0.03$, and $0.1$, each using distinct values of $\mu_c$ and $A$.
	}
	\label{fig:mu_of_I}
	\end{center}
\end{figure}

We first focus on the rheology in TXE, TXC, and pure shear flow.
Note that pure shear, like simple shear, is the only planar flow.
At each $\theta^{\dot{\epsilon}}$, the rheology seen in Fig. \ref{fig:mu_of_I} is characteristic of a monotonic $\mu(I)$ relationship where 
\begin{equation}
I = \dot{\epsilon}_S \langle d \rangle \sqrt{\rho/P}
\label{eq:in}
\end{equation}
is the inertial number, a scaled measure of the strain rate \cite{DaCruz2005, Jop2006}, and $\langle d \rangle = a$ is the average particle diameter.
With decreasing $I$, $\mu$ decreases and approaches a limiting value $\mu_c$ in the quasistatic limit.
With increasing interparticle friction $\mu_S$, $\mu$ reaches $\mu_c$ at a larger value of $I$.
$\mu$ is maximized in TXC and decreases with increasing $\theta^{\dot{\epsilon}}$ going to TXE.
The ratio of $\mu$ between the TXE and TXC limits, $\mu_\mathrm{TXE}/\mu_\mathrm{TXC}$, is known as the strength ratio $\Psi$ \cite{Fossum2006} and depends on both $\mu_S$ and $I$ as further discussed below. 
A subset of data was also generated for pressures of $10^{-4}$ and $10^{-6}$ but no significant change in $\mu(I)$ curves was observed reflecting the negligible pressure dependence in the hard-particle limit \cite{Jop2006, DeCoulomb2017, Srivastava2020b}.

For every $\theta^{\dot{\epsilon}}$ and $\mu_S$, the rise in $\mu$ with increasing $I$ is captured by a monotonic relation:
\begin{equation}
\mu - \mu_c = A I^\alpha
\label{eq:HB}
\end{equation}
where $\alpha$ and $A$ are fitted parameters.
This model is commonly used to describe dense granular rheology in simple shear \cite{Peyneau2008,DeGiuli2015, DeGiuli2016, DeCoulomb2017, Salerno2018, Srivastava2020b} but its efficacy across a wide range of Lode angles has not yet been tested.
To capture the change in $\Psi$ with $I$, there must be either a $\theta^{\dot{\epsilon}}$-dependent $\alpha$ or $A$.
Due to challenges in fitting power-law models \cite{Clemmer2021a}, our data is unable to rule out either option.
However, data is reasonably described using a $\theta^{\dot{\epsilon}}$-dependent value of $A$ and a $\theta^{\dot{\epsilon}}$-independent exponent of $\alpha = 0.4$, $0.7$, and $0.9$ for values of $\mu_S = 0.0$, $0.03$, and $0.1$, respectively.
An increase in $\alpha$ with $\mu_S$ has been previously identified in Refs.  \cite{DeGiuli2015,DeGiuli2016,DeCoulomb2017,Salerno2018, Srivastava2020b}.
At large $I > 0.1$, data begins deviating from Eq. \eqref{eq:HB} and may reflect a transition to a gas regime \cite{Jaeger1996}.

Despite being a rotational flow, results from simple shear simulations are also included in Fig. \ref{fig:mu_of_I} due to its common usage \cite{DaCruz2005, Peyneau2008, Salerno2018, Srivastava2020b, Singh2020}. 
Our simple shear data overlaps with results from Ref. \cite{Srivastava2020b} which used fully stress-controlled simulations and a Nos\'{e}-Hoover barostat.
In frictionless simulations, there is no significant difference between simple and pure shear. 
However for $\mu_S \ne 0$, simple shear curves are lower than pure shear. 
This is not totally unexpected as the kinematics of the flow differ subtly \cite{Giusteri2018}.
This behavior could be attributed to microstructural effects \cite{Srivastava2020b} although it is not explored here.
These two flow types offer the best chance of generalizing rheology to more complex models such as those found in Ref. \cite{Srivastava2020b}.
In fact, the simple shear model in Ref. \cite{Srivastava2020b} would reduce to a Reiner-Rivlin type model for pure shear due to the lack of rotation.

%%%%%%%%%%%%%%%%%%%%%%%%%%%%%%%%%%%%%%%%%%%%%%%%%%%%%%%%%%%%
%                        PSI
%%%%%%%%%%%%%%%%%%%%%%%%%%%%%%%%%%%%%%%%%%%%%%%%%%%%%%%%%%%%

\begin{figure}
	\begin{center}
	\includegraphics[width=0.48\textwidth]{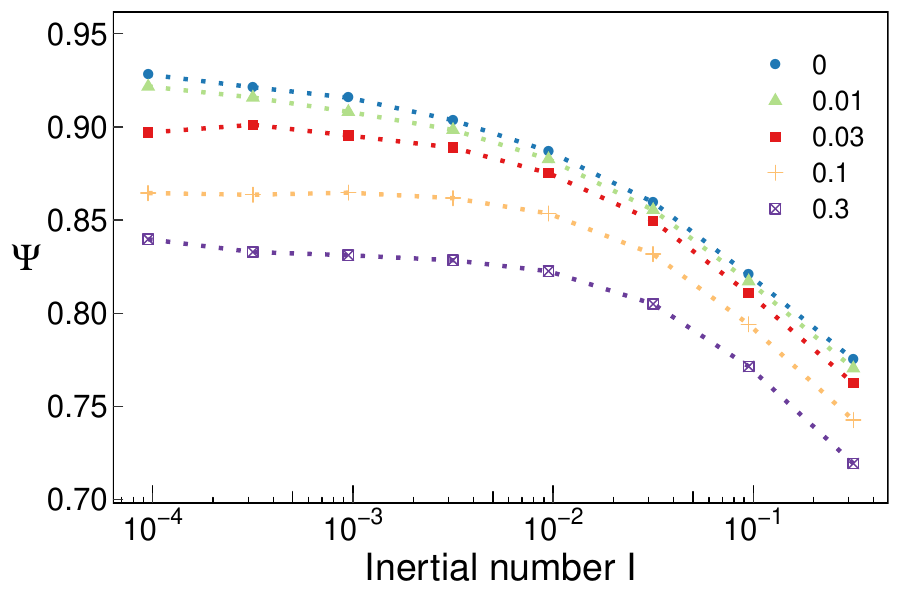}
	\caption{Strength ratio $\Psi$ as a function of $I$ for different $\mu_S$.} 
	\label{fig:Psi_v_I}
	\end{center}
\end{figure}

As mentioned above, $\Psi = \mu_\mathrm{TXE}/\mu_\mathrm{TXC}$ is a key measure of the effect of flow type on rheology and is plotted against $I$ in Fig. \ref{fig:Psi_v_I} for different $\mu_S$.
As $I$ increases, $\Psi$ decreases.
With increasing friction $\mu_S$, curves of $\Psi$ shift downward and saturate at larger $I$.
Crucially, this implies the stress envelope develops a stronger dependence on flow type at higher $I$ and $\mu_S$.
Interestingly at $\mu_S = 0$, curves reach a value of $\Psi \approx 0.93$. 
It is unknown to the authors whether a system could ever reach $\Psi = 1.0$ which would correspond to an isotropic rheology.
From our results, this could only be possible for frictionless systems near jamming as anisotropic effects otherwise emerge \cite{Giusteri2018,Srivastava2020b}.

%%%%%%%%%%%%%%%%%%%%%%%%%%%%%%%%%%%%%%%%%%%%%%%%%%%%%%%%%%%%
%                        CODIRECTIONALITY
%%%%%%%%%%%%%%%%%%%%%%%%%%%%%%%%%%%%%%%%%%%%%%%%%%%%%%%%%%%%

\begin{figure}
	\begin{center}
	\includegraphics[width=0.48\textwidth]{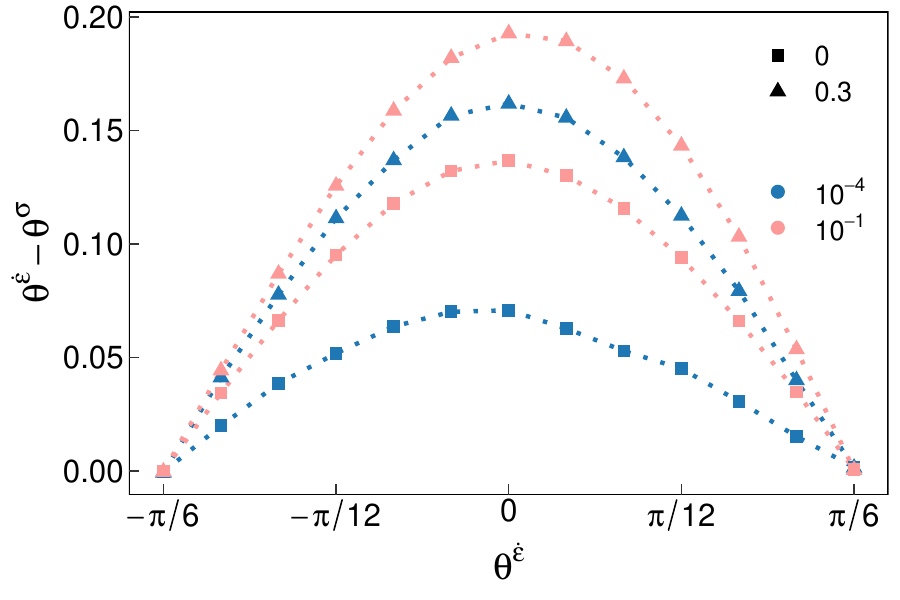}
	\caption{The difference in $\theta^{\dot{\epsilon}}$ and $\theta^{\sigma}$ as a function of $\theta^{\dot{\epsilon}}$ for the indicated values of $\mu_S$ (shape) and $I$ (color).
	}
	\label{fig:codirect}
	\end{center}
\end{figure}

As previously mentioned, the eigenvalues of the deviatoric stress and strain-rate tensors are not always proportional implying the tensors are not codirectional. 
The breakdown of codirectionality is greatest at intermediate $\theta^{\dot{\epsilon}}$ where $\theta^{\sigma}$ is smaller as seen in Fig. \ref{fig:codirect}.
The maximum deviation is around $4^\circ$ at $I = 10^{-3}$ and $\mu_S = 0$ but increases with increasing $I$ and $\mu_S$ up to $11^\circ$ at $I = 0.1$ and $\mu_S = 0.3$.
This effect originates from anisotropy in the contact network and the fabric tensor \cite{Thornton2010, Srivastava2020b} which is greater in frictional systems \cite{Cheal2018}.
In the TXE and TXC limits, this misalignment is minimized at all $I$ and $\mu_S$ and does not exceed $1^\circ$.
We therefore assume $\theta^\sigma$ and $\theta^{\dot{\epsilon}}$ are equivalent and the two tensors are codirectional in these limits, simplifying the following discussion of yield models.
This would imply $\sigma_{ij}$ and $\dot{\epsilon}_{ij}$ are simply related by a proportionality constant establishing two key limits for a fully tensorial model.

%%%%%%%%%%%%%%%%%%%%%%%%%%%%%%%%%%%%%%%%%%%%%%%%%%%%%%%%%%%%
%                        FITTING ENVELOPE
%%%%%%%%%%%%%%%%%%%%%%%%%%%%%%%%%%%%%%%%%%%%%%%%%%%%%%%%%%%%

Next we evaluate the shape of the steady-state envelope of the stress ratio: $\mu$ as a function of $\theta^{\sigma}$.
Two such envelopes at different values of $I$ are rendered in Fig. \ref{fig:example}.
The symmetry of the envelope reflects the equivalence of the three principal stresses.
At both $I$, $\mu$ is maximized in TXC and monotonically decreases with $\theta^{\sigma}$ before reaching a minimum at TXE.
As $I$ increases, not only does $\mu$ increase but the surface becomes more triangular.

The dependence of $\mu$ on $\theta^{\sigma}$ is often described using repurposed models of yield surfaces \cite{Phillips1965} which traditionally define the initial yielding at small strains rather than the stress envelope at constant values of $I > 0$ in steady-state flow.
In Mohr-Coulomb (MC) theory, a system flows if
\begin{equation}
(\sigma_1 - \sigma_3)/(\sigma_1 + \sigma_3) \ge \sin[\phi(I,\mu_S)]
\end{equation}
where $\phi$ is the angle of internal friction \cite{Panteghini2014}.
This model assumes the intermediate principal stress $\sigma_2$ is irrelevant producing a polygonal envelope with discontinuous derivatives at $\theta = \pm \pi/6$ (Fig. \ref{fig:example}).
The Drucker-Prager (DP) model alternatively assumes there is no dependence on the type of flow and only requires that $\mu$ exceeds a threshold $\mu_\mathrm{DP}(I,\mu_S)$, producing a circular profile (Fig. \ref{fig:example}) that is only correct if $\Psi = 1$.
From our results and simulations in Ref. \cite{Thornton2010}, neither model is accurate although they bound the actual response.
Note that cohesive terms in yield models were ignored for granular materials.

\begin{figure}
	\begin{center}
	\includegraphics[width=0.48\textwidth]{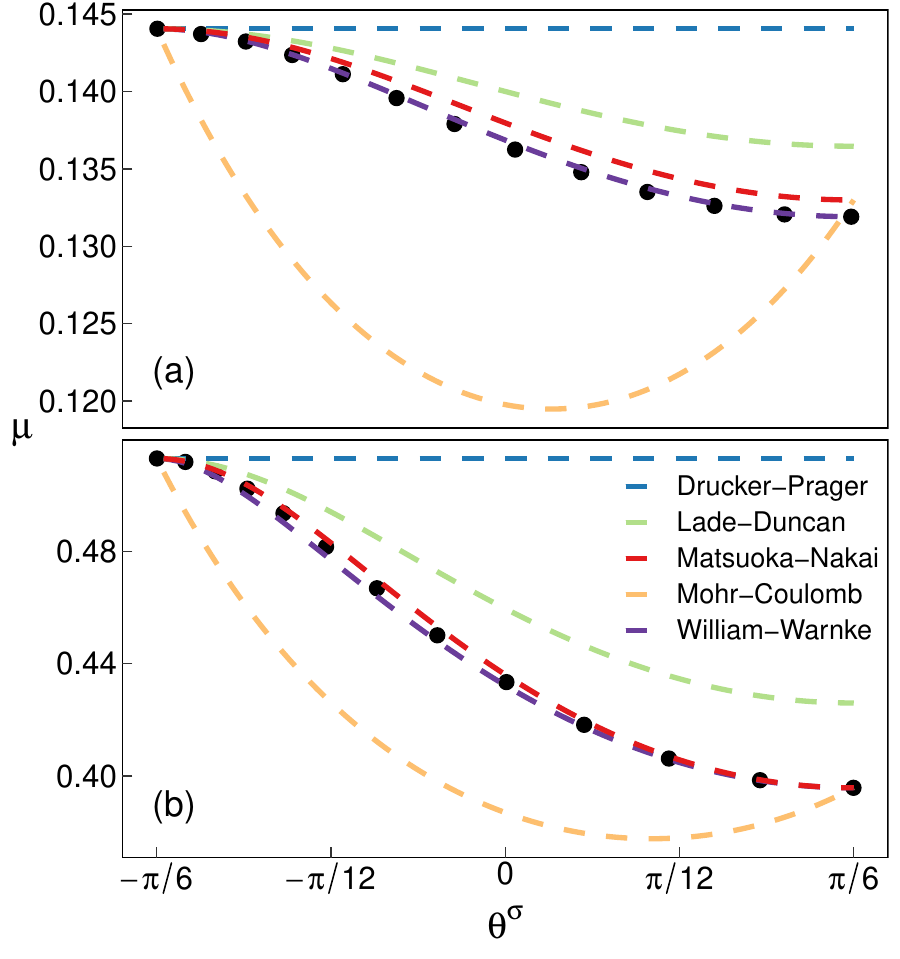}
	\caption{The shear stress as a function of $\theta^{\sigma}$ (data points)  and fitted models (lines) for (a) $I = 10^{-3}$ and $\mu_S = 0$ and (b) $I = 10^{-1}$ and $\mu_S = 0.3$.}
	\label{fig:mu_v_theta}
	\end{center}
\end{figure}

One segment of the envelope is plotted for $I = 10^{-3}$ and $\mu_S = 0.0$ in Fig. \ref{fig:mu_v_theta}(a) and $I = 10^{-1}$ and $\mu_S = 0.3$  in Fig. \ref{fig:mu_v_theta}(b). 
These two sets of data are chosen as they approximately maximize and minimize $\Psi$, respectively. 
Overlaid are DP and MC curves.
For DP, the critical stress ratio $\mu_\mathrm{DP}$ was simply set equal to $\mu_\mathrm{TXC}$.
For MC, $\mu_\mathrm{TXC}$ was used to calculate $\phi$ \cite{Lagioia2016}, although one could calculate $\phi$ in other flow types \cite{Fleischmann2020}.
As before, neither model is accurate but MC correctly predicts $\mu_\mathrm{TXE}$.

To account for this failure, more complex yield models have been devised including the William-Warnke (WW) \cite{William1975}, Matusoka-Nakai (MN) \cite{Matsuoka1974}, and Lade-Duncan (LD) \cite{Lade1975} models. 
The WW model interpolates between the DP and MC models using an elliptical function and is fit using values of $\mu_\mathrm{TXC}$ and $\Psi$ as described in Ref. \cite{Fossum2006}.
The MN and LD models use different combinations of stress invariants to construct yield criteria that include the effect of $\sigma_2$.
For the MN and LD models, we use a unified formulation from Ref. \cite{Lagioia2016} which is fit in terms of $\phi$ from the MC model.
Fits from these models are included in Fig. \ref{fig:mu_v_theta}. 

All models are constrained to predict $\mu_\mathrm{TXC}$ based on the fitting protocol while only the WW model is also constrained to predict $\mu_\mathrm{TXE}$, having been fit with two parameters.
Both the WW and MN models are fairly accurate while the LD model overpredicts $\mu_\mathrm{TXE}$.
Interestingly, this is the opposite of yielding where MN underpredicts and LD correctly predicts $\mu_\mathrm{TXE}$ \cite{Thornton2010, Huang2014,Fleischmann2014, Fleischmann2020}.
Although the WW and MN models both deviate from the data, they may be sufficiently accurate for continuum rheology models as the root-mean-square error across Lode angles is generally less than a few percentages of $\sigma_{TXC}$ for all $I$ and $\mu_S$ tested.

%%%%%%%%%%%%%%%%%%%%%%%%%%%%%%%%%%%%%%%%%%%%%%%%%%%%%%%%%%%%
%               DISCUSSION AND CONCLUSION
%%%%%%%%%%%%%%%%%%%%%%%%%%%%%%%%%%%%%%%%%%%%%%%%%%%%%%%%%%%%

In this work, DEM simulations were used to explore steady-state granular rheology over an extensive range of friction values, strain rates, and flow types.
A dataset of the deviatoric stress tensor is available in the Supplemental Material \footnote{See the Supplemental Material at the journal's website for a dataset of the average stress tensor measured at all friction coefficients, Lode angles, and inertial numbers considered in this study.}.
The type of flow has a significant effect on granular rheology, an effect that grows with increasing friction or inertial number.
The dependence on the type of flow is often simplified and described by Mohr-Coulomb or Drucker-Prager models although actual behavior lies in between the extremes of these two models.
There are other granular features which can also affect the stress envelope that should be studied such as other modes of interparticle friction \cite{Singh2020, Santos2020} and aspherical grain shapes \cite{Salerno2018}.
Finally, these results motivate the need to formulate tensorial rheological models which describe the effect of flow type \cite{Giusteri2018, Srivastava2020b, Coquand2021} and parametrize them using bulk simulations across flow types along with simulations of flows in complex geometries \cite{Silbert2001, Depken2007, Rycroft2009, Bhateja2020}.

\begin{acknowledgments}

I.S. acknowledges support from the U.S. Department of Energy, Office of Science, Office of Advanced Scientific Computing Research, Applied Mathematics Program under contract No. DE-AC02-05CH11231.
This work was performed at the Center for Integrated Nanotechnologies, a U.S. Department of Energy and Office of Basic Energy Sciences user facility.
Sandia National Laboratories is a multi-mission laboratory managed and operated by National Technology and Engineering Solutions of Sandia, LLC., a wholly owned subsidiary of Honeywell International, Inc., for the U.S. Department of Energy’s National Nuclear Security Administration under contract DE-NA-0003525.

\end{acknowledgments}

%%%%%%%%%%%%%%%%%%%%%%%%%%%%%%%%%%%%%%%%%%%%%%%%%%%%%%%%%%%%
%               REFERENCES
%%%%%%%%%%%%%%%%%%%%%%%%%%%%%%%%%%%%%%%%%%%%%%%%%%%%%%%%%%%%

%=====================================================================%
\newpage
\bibliographystyle{apsrev4-1}

\end{document}